\documentclass{iau} 
\usepackage{graphicx,natbib}
\usepackage{subfigure}

\title[Polarized emission of DSO/G2] 
{Detection of polarized continuum emission of the Dusty S-cluster Object (DSO/G2)}

\author[B. Shahzamanian et al.]   
{B. Shahzamanian$^1$, M. Zaja\v{c}ek$^{1,2}$, M. Valencia-S.$^1$, F. Peissker$^1$,\\A. Eckart$^{1,2}$, N. Sabha$^1$ \and M. Parsa$^{1,2}$}

\affiliation{$^1$Universit\"at zu K\"oln, Z\"ulpicher Strasse 77, D-50937 K\"oln, Germany\\ email: {\tt shahzaman, zajacek@ph1.uni-koeln.de} \\[\affilskip]
$^2$MPIfR, Auf dem H\"ugel 69, D-53121 Bonn, Germany}

\pubyear{2016}
\volume{322}  
\jname{The Multi-Messenger Astrophysics of the Galactic Centre}
\editors{S.N. Longmore, G. Bicknell \& R. Crocker, eds.}
\begin{document}

\maketitle

\begin{abstract}
A peculiar source in the Galactic center known as the Dusty S-cluster Object (DSO/G2) moves on a highly eccentric orbit around the supermassive black hole with the pericenter passage in the spring of 2014. Its nature has been uncertain mainly because of the lack of any information about its intrinsic geometry. For the first time, we use near-infrared polarimetric imaging data to obtain constraints about the geometrical properties of the DSO.
  We find out that DSO is an intrinsically polarized source, based on the significance analysis of polarization parameters, with the degree of the polarization of $\sim 30\%$ and an alternating polarization angle as it approaches the position of Sgr~A*. Since the DSO exhibits a near-infrared excess of $K_{\rm s}-L'>3$ and remains rather compact in emission-line maps, its main characteristics may be explained with the model of a pre-main-sequence star embedded in a non-spherical dusty envelope.  
\keywords{Galaxy: center, polarization, stars: pre--main-sequence}
\end{abstract}

\section{Dusty S-cluster Object (DSO/G2) and Near-infrared polarization}

\begin{figure}[!t]
  \centering
  \includegraphics[width=0.75\textwidth]{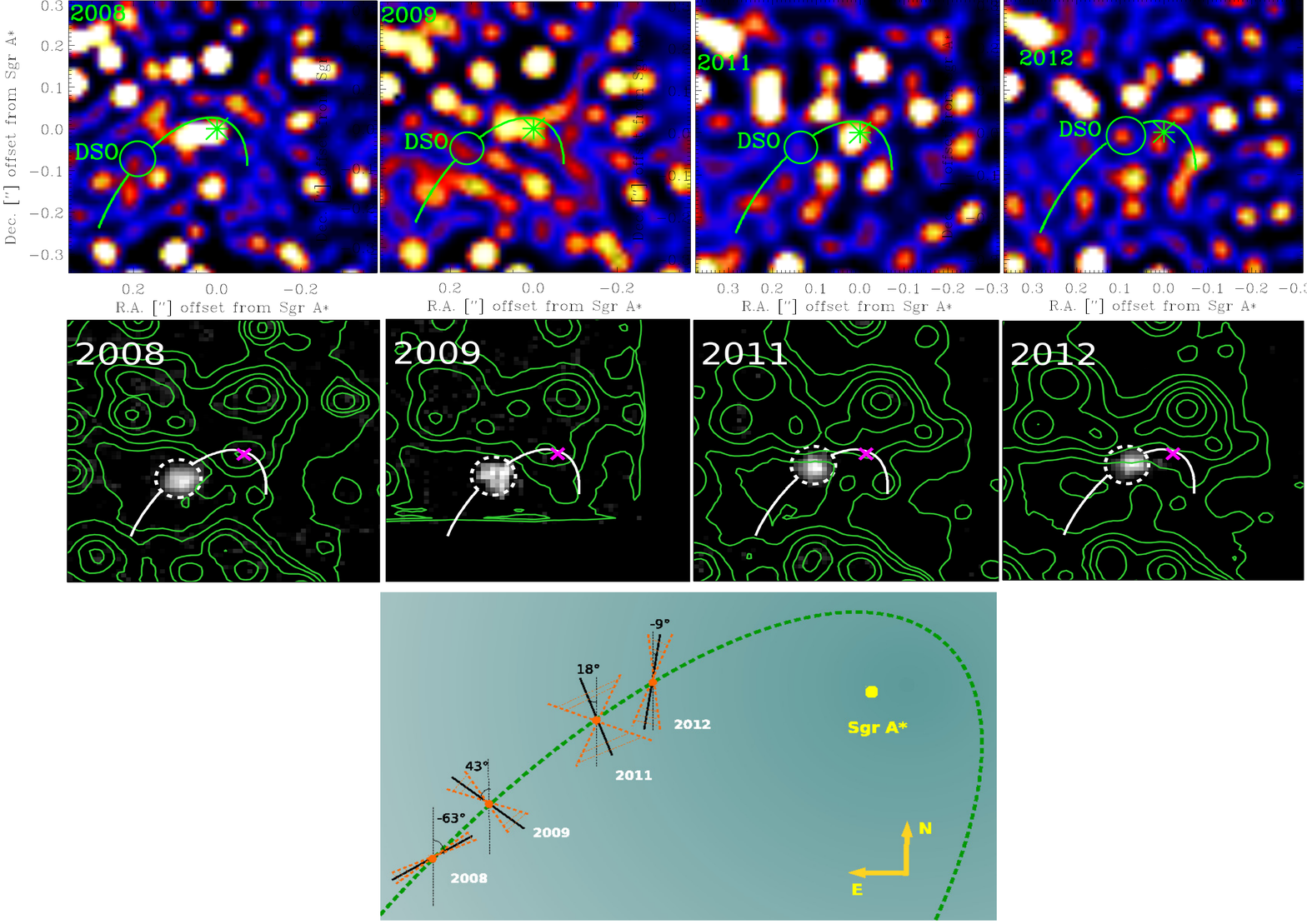}
  \caption{{\bf Top row:} Detection of the DSO continuum emission in $K_{\rm s}$ band in median deconvolved polarimetry images ($90^{\circ}$ channel of NACO imager) in different epochs 2008-2012. Bright spots correspond to stars. {\bf Middle row:} Br$\gamma$ line maps associated with the DSO for the same years in $0.8'' \times 0.8''$ SINFONI images. Contours correspond to the continuum emission of stars. {\bf Bottom row:} A sketch showing the DSO polarization angle variation.}
  \label{fig1}    
\end{figure}

The Dusty S-cluster object (DSO/G2), which was found in 2012 as a faint object approaching Sgr~A* \citep{2012Natur.481...51G}, can be primarily tracked in $L'$-band continuum and recombination line emission, mainly Br$\gamma$ in $K_{\rm s}$ band. So far DSO has remained compact in Br$\gamma$ line emission both before and after the periapse, see the analysis in \citet{2015ApJ...800..125V} and \citet{peissker16}. \cite{eckart2013} detected the first $K_{\rm s}$-band identification of this source in the continuum imaging data with a magnitude of ∼18.9. In Fig.~\ref{fig1} (middle row) we show the evolution of Br$\gamma$ emission in line maps for the epochs 2008-2012. Combined with the compact continuum $L'$-band emission \citep{2014ApJ...796L...8W}, the observations have shown that the DSO did neither stretch significantly nor disintegrate as was previously claimed \citep{2012Natur.481...51G,2015ApJ...798..111P}.

\vspace*{0.1cm}

In \citet{shahzamanian16} we use the near-infrared polarimetric imaging data to determine the polarization properties of the DSO for the first time. In addition to the study of continuum and line emissions from the DSO, the analysis of the $K_{\rm s}$ -band polarimetry data allows us to determine the properties of this source. The dust properties and the object geometries can be estimated quantitatively by the polarization analysis.  

\vspace*{0.1cm}

The $K_{\rm s}$-band polarimetry observations and the subsequent analysis and modelling (See also \citet{zajacek_iaus322} contribution in this volume) indicate the following results:

\begin{itemize}
 \item The DSO has a $K_{\rm s}$-band continuum counterpart with the flux density of $0.23\pm 0.04\,\rm{mJy}$, which corresponds to approximately $18.5$ magnitude.  
 \item The DSO is an intrinsically polarized source with the stable polarization degree of $\sim 30\%$, see Fig.~\ref{fig1}, top row. In different observing years (2008-2012), the measured polarization degree of this source is prominently higher than the foreground polarization which is 6.1\% in $K_{\rm s}$-band.
 \item The polarization degree stays approximately constant within uncertainties. The significance of the linear polarization measurements from Monte Carlo simulations is larger than $1 - 1/100\,000$.
 \item The obtained polarization angle of the DSO varies while the source moves towards the periapse (see Fig.~\ref{fig1}, bottom row). The change in the polarization angle is due to the internal influences (the circumstellar configuration) and/or the external influences (the source interaction with the accretion flow).
 \item The significant polarized emission as well as a large infrared excess may be explained by the model of a dust-enshrouded star that deviates from the spherical symmetry.
 \item The total and polarized flux density are matched by a composite stellar model consisting of the star, envelope, bipolar cavities, and the bow shock \citep{2014A&A...565A..17Z,2015arXiv150700237Z,2016MNRAS.455.1257Z}. 
\end{itemize}

\begin{thebibliography}{}



\bibitem[\protect\citeauthoryear{{Eckart}, A., {Mu{\v z}i{\'c}}, K., {Yazici}, S., {Sabha}, N., {Shahzamanian}, B., {Witzel}, G., {Moser}, L., {Garcia-Marin}, M., {Valencia-S.}, M., {Jalali}, B.{Bremer}}{{Eckart} et~al.}{2013}]{eckart2013}
{Eckart} A., {Mu{\v z}i{\'c}}, K., {Yazici}, S., et al. 2013, \textit{A\&A}, 551, A18




\bibitem[\protect\citeauthoryear{{Gillessen}, {Genzel}, {Fritz}, {Quataert},
  {Alig}, {Burkert}, {Cuadra}, {Eisenhauer}, {Pfuhl}, {Dodds-Eden}, {Gammie} \&
  {Ott}}{{Gillessen} et~al.}{2012}]{2012Natur.481...51G}
{Gillessen} S.,  {Genzel} R.,  {Fritz} T.~K., et al., \textit{Nature}, 481, 51


  
\bibitem[\protect\citeauthoryear{{Pfuhl}, {Gillessen}, {Eisenhauer}, {Genzel},
  {Plewa}, {Ott}, {Ballone}, {Schartmann}, {Burkert}, {Fritz}, {Sari},
  {Steinberg} \& {Madigan}}{{Pfuhl} et~al.}{2015}]{2015ApJ...798..111P}
{Pfuhl} O.,  {Gillessen} S.,  {Eisenhauer} F., et al.  2015, \textit{ApJ}, 798, 111

  \bibitem[\protect\citeauthoryear{{Peissker} \& {et
  al.}}{{Peissker} et~al.}{2016}]{peissker16}
{Peissker} F.,    {et al.} 2016, in preparation

\bibitem[\protect\citeauthoryear{{Shahzamanian}, {Eckart}, {Zaja\v{c}ek} \& {et
  al.}}{{Shahzamanian} et~al.}{2016}]{shahzamanian16}
{Shahzamanian} B., {Eckart} A., {Zaja\v{c}ek} M., {et al.} 2016, \textit{A\&A} in print, ArXiv 1607.04568



\bibitem[\protect\citeauthoryear{{Valencia-S.}, {Eckart}, {Zaja{\v c}ek} \& {et
  al.}}{{Valencia-S.} et~al.}{2015}]{2015ApJ...800..125V}
{Valencia-S.} M.,  {Eckart} A.,  {Zaja{\v c}ek} M.,    {et al.} 2015, \textit{ApJ},
  800, 125
  
  \bibitem[\protect\citeauthoryear{{Witzel}, {Ghez}, {Morris}, {Sitarski},
  {Boehle}, {Naoz}, {Campbell}, {Becklin}, {Canalizo}, {Chappell}, {Do}, {Lu},
  {Matthews}, {Meyer}, {Stockton}, {Wizinowich} \& {Yelda}}{{Witzel}
  et~al.}{2014}]{2014ApJ...796L...8W}
{Witzel} G.,  {Ghez} A.~M.,  {Morris} M.~R., et al.,  2014, \textit{ApJL}, 796, L8

\bibitem[\protect\citeauthoryear{{Zaja{\v c}ek}, {Karas} \& {Eckart}}{{Zaja{\v
  c}ek} et~al.}{2014}]{2014A&A...565A..17Z}
{Zaja{\v c}ek} M.,  {Karas} V.,    {Eckart} A.,  2014, \textit{A\&A}, 565, A17

\bibitem[\protect\citeauthoryear{{Zaja\v{c}ek}, {Eckart}, {Peissker}, {Karssen} \&
  {Karas}}{{Zaja\v{c}ek} et~al.}{2015}]{2015arXiv150700237Z}
{Zaja\v{c}ek} M.,  {Eckart} A.,  {Peissker} F., et al.
  2015, Proceedings of 24th WDS, ArXiv 1507.00237
  
  \bibitem[{{Zaja{\v c}ek} {et~al}\mbox{.}(2016){Zaja{\v c}ek}, {Eckart},
  {Karas}, {Kunneriath}, {Shahzamanian}, {Sabha}, {Mu{\v z}i{\'c}}, \&
  {Valencia-S.}}]{2016MNRAS.455.1257Z}
{Zaja{\v c}ek} M., {Eckart} A., {Karas} V., et al., 2016, \textit{MNRAS}, 455, 1257

\bibitem[\protect\citeauthoryear{{Zaja\v{c}ek} et~al.}{}]{zajacek_iaus322}
{Zaja\v{c}ek} M.,  {Shahzamanian} B.,  {Valencia-S.} M., et al.,
  Proceedings of IAUS 322, this volume



\end{thebibliography}
\end{document}